\def\today{\ifcase\month\or
  January\or February\or March\or April\or May\or June\or
  July\or August\or September\or October\or November\or December\fi
  \space\number\day, \number\year}

\documentstyle[prl,twocolumn,aps]{revtex}
\def\baselinestretch{1.0}
\begin{document}
\draft

\title{Resonant Tunneling through Multi-Level and Double Quantum Dots}

\author{T. Pohjola$^{1,2}$, J. K\"onig$^2$, M. M. Salomaa$^1$, 
J. Schmid$^3$, H. Schoeller$^2$, and Gerd Sch\"on$^{1,2}$}

\address{$^1$Materials Physics Laboratory, 
Helsinki University of Technology, 02150 Espoo, Finland\\
$^2$Institut f\"ur Theoretische Festk\"orperphysik, Universit\"at Karlsruhe,
76128 Karlsruhe, Germany\\
$^3$ Max-Planck-Institut f\"ur Festk\"orperforschung,
70569 Stuttgart, Germany}

\date{\today}

\maketitle

\begin{abstract}

We study resonant tunneling through
quantum-dot systems in the presence of strong Coulomb repulsion 
and coupling to the metallic leads. 
Motivated by recent experiments we concentrate on
(i) a single dot with two energy levels and (ii) a double dot with  one
level in each dot. Each level is twofold spin-degenerate. Depending
on the level spacing these systems are physical realizations of 
different Kondo-type models.
Using a real-time diagrammatic formulation we evaluate the 
spectral density and the non-linear conductance. The latter shows  
a novel triple-peak resonant structure.

\end{abstract}
\pacs{73.40.Gk, 72.15.Qm, 73.20.Dx, 73.50.Fq}

\narrowtext

Quantum transport of electrons through discrete energy levels in quantum
dots has been studied both experimentally \cite{reviews,kastner}
and  theoretically
\cite{HS-Hab,klassisch,beenakker,meir,weinmann,bruder}.
In small dots the Coulomb repulsion is strong and may 
yield the dominant energy scale, which gives rise to
 Coulomb blockade phenomena 
(see e.g. Refs.~\cite{klassisch,beenakker,meir,weinmann,bruder} 
for single dots and
Refs.~\cite{doppeldot,stafford,pals,golden} for multiple dots).
At high temperatures, or when the coupling between the dots and 
metallic leads is weak, the electrons tunnel through the 
system sequentially. 
At low temperatures and for strong coupling to the leads resonant 
tunneling processes contribute significantly to the current
\cite{HS-Hab,Kondo,glazman2,KSS-PRL,Her-Dav-Wil,Mei-Win-Lee}.

The purpose of this Letter is to investigate electronic transport phenomena
through two ultrasmall quantum-dot systems:
(i) a single dot with two energy levels and 
(ii) a double dot with a single level in each dot (see 
Fig.~\ref{fg:Double dot}). 
We will show that the two models can be mapped onto each other.
Thus, it is sufficient to concentrate on one, the single-dot
model with two levels. 
In both cases we assume strong Coulomb repulsion between the 
electrons, such that we have to consider only two charge
states, e.g. the empty or the singly-occupied dot system. 
We further allow for strong coupling between the quantum dot(s) and the 
metallic leads adjacent to them. 
This gives rise to resonant tunneling and -- as has been shown for a single 
dot with one energy level \cite{KSS-PRL,Her-Dav-Wil,Mei-Win-Lee}
-- nonequilibrium Kondo effects. 
In the following we consider the case with vanishing magnetic field. 
Accordingly the single-electron states are  spin-degenerate, which is 
a prerequisite for Kondo effects.
We suggest that the models could be realized experimentally in a setup
resembling those in Refs.
\cite{Ralph&Buhrman,Ral-etal,Taruchaetal,Westerveltetal}.

A systematic description of resonant tunneling phenomena has been 
developed in Refs.~\cite{HS-Hab,KSS-PRL,method} in terms of
a real-time diagrammatic technique. It has been applied to the 
problems of electron transport through a single-level quantum dot as well 
as through a small metal island, both 
connected to reservoirs via tunnel junctions. 
In the quantum dot Kondo-type effects 
manifest themselves as a zero-bias 
maximum in the differential conductance.
Furthermore, a zero-bias minimum has been predicted for the case
where the level lies above the electrochemical 
potentials of the reservoirs \cite{KSS-PRL}.
Generalizing this method we 
calculate the spectral density of the two-level dot and the
differential conductance for the current through the dot. 
As novel result we find a multiply peaked resonant structure both in the
spectral functions and the differential conductance. 
When the two levels in the dot are positioned below the electrochemical 
potentials of the leads a triple-peak structure emerges in the nonlinear
conductance: a zero-bias maximum with one satellite peak on each side. 
When the levels lie above the electrochemical potentials the peaks turn into 
notches.
Depending on the level spacing our model interpolates between an 
$S=1/2$ and an $S=3/2$ Kondo-type model (generalized to include transitions
between all 'spin states').

We denote the two energy levels of the dot by
$\varepsilon_+$ and $\varepsilon_-$
(the choice of $+$ and $-$ arises from the eigenstates of the 
double dot system). 
Both levels are spin-degenerate, 
$\varepsilon_{i\uparrow}=\varepsilon_{i\downarrow}=\varepsilon_i$ 
for $i\in\{+,-\}$.
The Hamiltonian is a generalization of the Anderson model,
${\cal H}={\cal H}_{0,{\rm D}}+{\cal H}_{0,{\rm res}} +{\cal H}_{\rm T}$.
The term ${\cal H}_{0,{\rm D}}=\sum_{i\sigma} 
\varepsilon_{i\sigma} c_{i\sigma}^\dag c_{i\sigma}^{} 
+ U_0 n_{\rm D}(n_{\rm D}-1)$ 
describes the isolated quantum dot.
The interaction of the electrons in the dot is accounted for by $U_0$,
with $n_{\rm D}=\sum_{i\sigma}c_{i\sigma}^\dag c_{i\sigma}^{}$ being the
electron number. 
The electrons in the reservoirs are taken to be non-interacting,
${\cal H}_{0,{\rm res}}=\sum_{\alpha\in \{\rm R,L\}} 
\sum_{k\sigma} \varepsilon_{\alpha k\sigma} 
a_{\alpha k\sigma}^\dag a_{\alpha k\sigma}^{}$.
The tunneling Hamiltonian 
${\cal H}_{\rm T}=\sum_{\alpha ki\sigma}\bigl( T_{ki\sigma}^\alpha 
c_{i\sigma}^\dag a_{\alpha k\sigma}^{} + {\rm h.c.}\bigr)$
describes tunneling between the dot and the reservoirs.

The double-dot system  coupled in series between two reservoirs 
as shown in Fig.~\ref{fg:Double dot} with one spin-degenerate level in each  
dot is described by the Hamiltonian
${\cal H}={\cal H}_{0,{\rm D}}+{\cal H}_{\rm 0,res}+{\cal H}_{\rm t}+{\cal
H}_{\rm T}$. 
The Hamiltonian of the two capacitively coupled dots is 
${\cal H}_{0,{\rm D}}=\sum_{d\sigma} \varepsilon_{d\sigma} c_{d\sigma}^\dag 
c_{d\sigma}^{} + U_{12}n_{\rm l}n_{\rm r} + U_0 \sum_d
n_{d\uparrow}n_{d\downarrow}$. In addition to the
terms referring to both dots it contains an additional interaction term 
$U_{12}n_{\rm l}n_{\rm r}$. 
The index $d\in\{{\rm l,r}\}$ denotes the left or right dot, respectively. 
The term ${\cal H}_{\rm t}=\sum_{\sigma}
\bigl( t\,c_{{\rm l}\sigma}^\dag c_{{\rm r}\sigma}+
{\rm h.c.}\bigr)$ 
describes the tunneling between the dots, while the 
tunneling Hamiltonian ${\cal H}_{\rm T}$ is modified 
such that it only couples each dot to the lead adjacent to it.

The double-dot model can be mapped onto the two-level single-dot model
in the regime of large charging energy, i.e., when $U_0$ and $U_{12}$ 
are larger than all other energy scales of the system.
Then it is sufficient to consider only two adjacent charge states of the 
double dot. For appropriate values of a gate voltage coupled to the dots
these states are the empty and the singly charged dot system.
Then, we  first solve the exact eigenstates of the double 
dot isolated
from the leads, i.e., the eigenstates of $H_{\rm 0,D}+H_t$. They  are 
symmetric and antisymmetric combination of the single-dot states in the left 
and right-hand dots 
$\vert +\rangle=\alpha\vert l\rangle+\beta\vert r\rangle$ 
and $\vert -\rangle=\beta\vert l\rangle-\alpha\vert r\rangle$, 
with energies $\varepsilon_+$ and $\varepsilon_-$, 
respectively. 
The tunneling ${\cal H}_{\rm T}$ couples the leads to these new eigenstates.
Thus the effective tunneling matrix elements are
$ T_{\rm L,+} = \alpha T \;,\; T_{\rm R,+} 
= \beta T \;,\; T_{\rm L,-} = \beta T \;,
\; T_{\rm R,-} = -\alpha T $.
Except for this modification of the matrix elements the
double-dot system coincides with the two-level single dot.

The mapping is not possible if we consider doubly-occupied states.
The interaction between electrons in the same dot is
stronger than between electrons in different dots. 
Hence two electrons in a double dot prefer to stay 
in different dots, and the low energy two-electron states are 
formed from these single-electron states only. 
In contrast in a single dot the two-electron states are
composed of all single-electron states.
By symmetry, in the case of  three and four electron states 
the mapping is possible again.

Physical quantities of interest can be expressed as quantum statistical
averages. 
In particular, the d.c. current through the quantum dot takes the form
\begin{equation}
        I=I_{\rm L}=-I_{\rm R}=
        -{ie\over\hbar}\sum_{ki\sigma}\Bigl(T_{ki\sigma}^{L}
        \langle a_{\alpha k\sigma}^\dag c_{i\sigma}\rangle -{\rm H.c.}\Bigr).
\end{equation}
The statistical average of an arbitrary operator $O(t)$ can, quite generally,
be expressed as
\begin{equation}\label{eq:Average}
        \langle O(t)\rangle ={\rm tr}\left\{\rho_0 T_{\rm K} \exp \left(
        -i\int_{\rm K} dt' {\cal H}_{\rm T} (t')_{\rm I} \right) 
        O(t)_{\rm I}\right\}.
\end{equation}
The time integration is performed along the {\it Keldysh contour}, i.e.,
forward in time from some initial time $t_{\rm i}$ to $t$ 
and then backward from $t$ to $t_{\rm i}$.
The time-ordering operator $T_{\rm K}$ arranges 
operators with respect to this 
Keldysh contour.
The subscript {\rm I} denotes the interaction picture with respect to 
${\cal H}_{\rm 0,D}+{\cal H}_{\rm 0,res}$.

The leads are assumed to be large and in thermal equilibrium.
Their electrochemical potentials are kept fixed at 
$\mu_{\rm r}=-eV_{\rm r}$ (for r=L,R),
and their electron distributions are described with Fermi functions.
We are interested in  the time evolution of 
the quantum states of the dot system, and 
therefore consider the reduced density matrix of the dot. 
To do this we first expand the exponential part in Eq.~(\ref{eq:Average})
into a series of the form
\begin{eqnarray}\label{eq:Expansion}
        \lefteqn{
        T_{\rm K}\: \exp \left( -i\int_{\rm K} dt'\:
        {\cal H}_{\rm T}(t')_{\rm I} \right)
        =\sum_{m=0}^\infty (-i)^m\cdot}\\
        & &{\int_{\rm K}\!dt_1\int_{\rm K}\!dt_2\cdots\!
        \int_{\rm K}\!dt_m} T_{\rm K} \! 
        \left\{{\cal H}_{\rm T}(t_1)_{\rm I} 
        {\cal H}_{\rm T}(t_2)_{\rm I} \cdots {\cal H}_{\rm T}(t_m)_{\rm I}
        \right\}.\nonumber 
\end{eqnarray}
Here the times of the integrals are ordered with respect to the Keldysh 
contour, such that $t_1>t_2>\cdots >t_m$. 
The expectation value of this series is taken by tracing over the 
non-interacting reservoirs in the density matrices.
We can perform the trace over the reservoir degrees of freedom exactly
since the unperturbed Hamiltonian ${\cal H}_{\rm 0,res}$ is bilinear 
in the lead-electron operators. 
These facts allow us to apply Wick's theorem, and the forward and backward 
propagators become coupled due to the trace over the leads. 
In Refs.~\cite{HS-Hab,KSS-PRL,method} a systematic diagrammatic 
technique has been developed for presenting the terms of the expansion 
Eq.~(\ref{eq:Expansion}). 
There are well-defined rules for evaluating an arbitrary diagram and it is 
possible to identify and study processes of different orders.

The current can be expressed in terms of the spectral densities
$A_{i\sigma}(\omega)$ of the electron states in the dot as \cite{Meir&Wingreen}
\begin{eqnarray}\label{eq:The current}
        \lefteqn{\!\!\!\!\!\!\!\!\!\!\!\!\!\!\!\!\!\!\!\!\!\!\!\!\!\!\!\!\!\!\!
        I = \sum_{i,\sigma}{e\over\hbar}\int_{-\infty}^\infty d\omega
        {\Gamma_{i\sigma}^R\Gamma_{i\sigma}^L\over 
        \Gamma_{i\sigma}^R+\Gamma_{i\sigma}^L}\cdot}\nonumber\\
        \;\;\;\;\;\;\;\;\;\;\;
        &&\left[f(\omega-\mu_{\rm R})-f(\omega-\mu_{\rm L})\right]
        A_{i\sigma}(\omega).
\end{eqnarray}
The factors $\Gamma_{i\sigma}^{\alpha}=2\pi\sum_k\left|
T_{ki\sigma}^{\alpha}\right|^2\delta(E-\varepsilon_{\alpha k\sigma})$
describe the coupling between the dot level $i$ and the lead $\alpha$ 
and they are assumed to be independent of energy \cite{com}.
The spectral function describes the spectrum of possible excitation of the
system.
In general there exist separate Green's and spectral 
functions for the four states of the dot: one for each spin of each level.
Here we restrict ourselves to the zero-field case such that the
levels and the corresponding spectral functions are spin-degenerate.

In general, we can not sum up {\it all} possible contribution from the 
expansion Eq.~(\ref{eq:Expansion}).
Similar as in Ref.~\cite{HS-Hab,Kondo,KSS-PRL} we proceed in a conserving 
approximation, taking into account non-diagonal matrix elements of the total 
density matrix up to the difference of one electron-hole pair 
excitation in the leads. In this case we find a set of integral equations
for the spectral densities $A_{i\sigma}(\omega)$.
While the  coupling between the dot and the reservoirs
is described within this
approximation scheme, the double-dot states are solved exactly.
This is reasonable when the coupling strength between the dots is 
at least comparable to the coupling between the dots and the leads.

In the following, we choose a symmetrical bias 
$\mu_{\rm L}=-\mu_{\rm R}=eV/2$ and consider equal-strength coupling of 
the dot to the leads $\Gamma_{\rm L}=\Gamma_{\rm R}$.
The resulting conductance curves are symmetrical as a function of the voltage.
The values for the energy levels that we mention, 
e.g., in the context of the illustrations, refer to the level 
positions before the renormalization due to tunneling. 
However, when we talk about levels being above or below the electrochemical
potentials we refer to the renormalized measurable level positions. 
Finally, we choose the temperature $k_{\rm B}T=0.02\Gamma$, 
where $\Gamma=\Gamma_{\rm L}+\Gamma_{\rm R}$ provides the energy scale,  
such that the thermal energies are always lower than the level spacing 
in the dots and much lower than the charging energy.

In the spectral densities $A_i(\omega)$ there appear a complex, 
energy-dependent self-energy term which yields a broadening as well as 
renormalization of the dot levels.
The energy-dependence of the self-energy term is due to the 
resonant-tunneling processes,
 and it results in a non-trivial functional form of the spectral functions.
On top of this structure  Kondo-type resonances  may appear as a sharp 
peaked structure at the position(s) of the electrochemical potential(s). 
These general effects of resonant tunneling have been found already in the 
investigations of a single one-level quantum dot
\cite{KSS-PRL,Mei-Win-Lee}.

We keep the lower level below the electrochemical potentials of the 
leads such that $\varepsilon_+\lesssim-\Gamma$.
Tuning the upper level $\varepsilon_-=\varepsilon_++\Delta\varepsilon$ 
allows a crossover from the one-level limit, 
$\varepsilon_-\rightarrow+\infty$, 
to a fourfold-degenerate level, $\varepsilon_+=\varepsilon_-$. 
For the single-dot model the level separation $\Delta\varepsilon$ reflects 
the actual physical level separation, whereas it is proportional to 
the coupling strength between the two dots for the double-dot model.
In the limit of infinite level separation the spectral density 
$A_+(\omega)$ reproduces results obtained 
before for a single level~\cite{KSS-PRL,Mei-Win-Lee}.
These are the level renormalization and broadening, as well as 
-- for $\varepsilon_+\lesssim-\Gamma$ --  Kondo-type resonances
at the electrochemical potentials of the leads. 
For a low lying level this corresponds to the
 $S=1/2$ Kondo problem, however generalized to a nonequilibrium 
situation if a transport voltage is applied. 

When the second level is included, the Kondo peak of the single-level 
model persists. However, the renormalization effect is stronger and, 
as long as $\varepsilon_+,\varepsilon_-\lesssim-\Gamma$,
a further peak in the spectral functions appears (see the 
dashed curves in Fig.~\ref{fg:Split}).
This peak is shifted from the electrochemical potentials by 
$\Delta\varepsilon$ below (above) the Kondo peak in the spectral 
function for the lower (upper) level. 
In the other extreme of zero level separation we have a fourfold degenerate 
case,  and the model corresponds approximately 
to a spin $S=3/2$ single-channel Kondo model (in the limit of low
lying levels, furthermore transitions between all 'spin states' are
allowed in the present case).
In this case the resulting spectral functions are equal to each 
other and have a much more pronounced Kondo peak 
(see the solid curve in Fig.~\ref{fg:Split}) 
than in the single level case.

The large resonant peak, seen when $\varepsilon_+=\varepsilon_-$, can be 
viewed as two Kondo peaks being on top of each other. 
When the degeneracy is lifted  the peak splits into two,
separated by $\Delta \epsilon$. One peak remains pinned at zero since 
the spin-degeneracy of the levels continues to lead to a Kondo peak.
This result differs from the situation where the spin degeneracy of a
single level is lifted by a magnetic field~\cite{KSS-PRL}.
When the upper level is lifted far above the lower one the 
properties of the system approach those of a single-level 
dot thus exhibiting just one Kondo peak.

The spectral functions are normalized. 
Since we have neglected multiple 
occupancy, the normalization is
\begin{equation}
        \int_{-\infty}^\infty d\omega\;A_{i\sigma}(\omega)
        =p_0+p_{i\sigma}\;,
\end{equation}
where $p_0$ and $p_{i\sigma}$ are the probabilities of an empty dot
and that with single electron state $i\sigma$ occupied.
This expression allows us to relate
differences in the magnitudes of the spectral functions
to differences in occupation probabilities,  
e.g. when the spectral functions corresponding to different
levels are compared to each other.

The bias voltage has the effect of splitting all the resonant peaks in 
the spectral functions by $eV$, see Fig.~\ref{fg:A with eV}. 
The peaked structure of the spectral functions is reflected in a novel 
structure in the  nonlinear conductance, see Fig.~\ref{fg:Conductance}. 
Instead of a single zero-bias peak, known from the single level dot,
there appear now three sharp resonant 
peaks between the broad peaks corresponding to the renormalized
energy levels. 
The zero-bias anomaly, characteristic for the single level case remains, 
but two additional satellite peaks appear at $eV=\pm\Delta\varepsilon$.
When the level separation is increased the new peaks decrease in size
while the peak at zero bias remains unchanged.
When the levels lie above the electrochemical potentials, the 
nonlinear conductance exhibits a triplet of minima with 
one minimum at the electrochemical potentials and two 
satellites on each side. 
When the levels are on opposite sides of the electrochemical 
potentials the spectral functions as well as the conductance show very 
varied structures, which are not easily classified.

The systems considered here have not yet been investigated experimentally.
Ralph and Buhrman \cite{Ralph&Buhrman} have measured
tunneling -- enhanced by the Kondo effect --
through a localized state of an impurity equivalent to
a single one-level dot.
Some unexplained features in their results
may be due to the existence of the other levels above the 
electrochemical potentials. 
The model of a single- or  two-level dot with only one electron in it
could possibly be realized with the setup used by 
Tarucha {\it et al.}~\cite{Taruchaetal}. 
They argue that they can resolve situations with one or few
one spin-degenerate levels and one or few electrons in the dot. 
In particular, they report that a third electron inserted into the dot 
finds two spin-degenerate levels whose separation can be continuously 
varied by a magnetic field. 
In this system, if the gate voltage is fixed near the 
first or third conductance peak the single- or 
double-level dot is realized, respectively.
The difference should be observable in the nonlinear conductance
provided that the zero-bias anomalies can be resolved.
The relevant parameters reported in Ref.~\cite{Taruchaetal}
-- the coupling strength (tunneling rate) $\Gamma$, voltage, and 
temperature -- are already close to those used in our calculations.

In conclusion, we have described
 electronic transport through a two-level quantum 
dot and a single-level double dot coupled between reservoirs. 
A real-time diagrammatic technique provides a very general tool 
for the description of various systems, both in the perturbative and 
nonperturbative regime. 
We have found a novel triple-peak structure in the 
nonlinear conductance of these 
systems and suggest experiments where it is observable.

\section*{Acknowledgments}

We are grateful to L.P. Kouwenhoven, M. Paalanen, and J. Pekola
for discussions. This work has been supported by the ``Deutsche 
Forschungsgemeinschaft'' as part of ``Sonderforschungbereich 195.''
One of us (GS) also gratefully acknowledges the support 
through an A.~v.~Humboldt Research Award of the Academy of Finland.

\begin{figure}
\caption{Double dot with one level with energy $\varepsilon_i$, 
        in each dot.
        The tunneling matrix elements of the barriers are $T_{\rm L,l}$ and 
        $T_{\rm R,r}$ for tunneling between the dots and leads and $t_{l,r}$ 
        for tunneling between the dots.}
\label{fg:Double dot}
\end{figure}

\begin{figure}
\caption{Spectral function $A_{+(\sigma)}(\omega)$ with 
        $\varepsilon_+=-2.0\,\Gamma$ and 
        $\varepsilon_-=\varepsilon_+ +\Delta\varepsilon$.
        The large Kondo peak of the four-fold degenerate level splits when 
        the degeneracy is lifted. 
        The separation equals the level spacing,
        $\Delta\varepsilon\in\{0,0.2\Gamma,0.4\Gamma\}$.}
\label{fg:Split}
\end{figure}

\begin{figure}
\caption{Spectral functions $A_+(\omega)$ when 
        a bias voltage $V$ is applied.
        The voltage splits all the resonant peaks by 
        energy $eV$ and the peaks
        cross each other when $eV=\Delta\varepsilon$.}
\label{fg:A with eV}
\end{figure}

\begin{figure}
\caption{The figure shows apart from the broad main peak in the non-linear
        conductance, which is sometimes referred to as the resonant 
        peak, sharper resonant peaks close to zero. 
        The inset shows the same enlarged. 
        The solid curve is for the four-fold degenerate level.
        The dashed curves are for split levels,
        $\Delta\varepsilon =0.2\Gamma$ (long dashed) and 
        $\Delta\varepsilon = 0.4\Gamma$ (short dashed).
        The conductance $g(V)$ is expressed in units of $e^2/h$.}
\label{fg:Conductance}
\end{figure}


\begin{references}
\bibitem{com}
The quantities
 $\Gamma_{ij\sigma}^{\alpha}= 2\pi\sum_k
T_{ki\sigma}^{\alpha *} T_{kj\sigma}^{\alpha}
\delta(E-\varepsilon_{\alpha k\sigma})$ are assumed to be diagonal 
with respect to $i,j$.
\end{references}
\end{document}